%% file: main.tex
\documentclass[%
 reprint,
 amsmath,amssymb,
 aps,
]{revtex4-2}

\usepackage{graphicx}% Include figure files
\usepackage{dcolumn}% Align table columns on decimal point
\usepackage{bm}% bold math
\begin{document}

\preprint{APS/123-QED}

\title{Experimental loopback boson sampling}
\author{Yu.A. Biriukov}
\altaffiliation[e-mail: ]{biriukov.ia18@physics.msu.ru}
\affiliation{
 Quantum Technology Centre and Faculty of Physics, M.V. Lomonosov Moscow State University, 1 Leninskie Gory, Moscow 119991, Russia
}
\author{R.D. Morozov}
\affiliation{
 Quantum Technology Centre and Faculty of Physics, M.V. Lomonosov Moscow State University, 1 Leninskie Gory, Moscow 119991, Russia
}
\author{K.I. Okhlopkov}
\affiliation{
Russian Quantum Center, 30 Bolshoy bul'var building 1, Moscow 121205, Russia
}
\author{I.V. Dyakonov}
\author{N.N. Skryabin}
\affiliation{
 Quantum Technology Centre and Faculty of Physics, M.V. Lomonosov Moscow State University, 1 Leninskie Gory, Moscow 119991, Russia
}
\affiliation{
Russian Quantum Center, 30 Bolshoy bul'var building 1, Moscow 121205, Russia
}
\author{M.V. Rakhlin}
\author{A.I. Galimov}
\author{G.V. Klimko}
\author{ S.V. Sorokin }
\author{ I.V. Sedova}
\author{ M.M. Kulagina}
\author{ Yu.M. Zadiranov}
\author{ A.A. Toropov}
\affiliation{
Ioffe Institute, 26 Polytekhnicheskaya street, St. Petersburg 194021, Russia
}
\author{S.A. Zhuravitskii}
\author{M.A. Dryazgov}
\author{ K.V. Taratorin}
\author{A.A. Korneev}
\author{S.P. Kulik}
\affiliation{
 Quantum Technology Centre and Faculty of Physics, M.V. Lomonosov Moscow State University, 1 Leninskie Gory, Moscow 119991, Russia
}
\author{S.S. Straupe}
\affiliation{
 Quantum Technology Centre and Faculty of Physics, M.V. Lomonosov Moscow State University, 1 Leninskie Gory, Moscow 119991, Russia
}
\affiliation{
Russian Quantum Center, 30 Bolshoy bul'var building 1, Moscow 121205, Russia
}

\date{\today}% It is always \today, today,

\begin{abstract}
We present an experimental demonstration of boson sampling enhanced by optical feedback lines, a novel approach that introduces temporal correlations among photons to amplify computational complexity. We utilize a 25-mode femtosecond laser-written interferometer with five output channels connected to five input channels to create correlations between consecutive photon arrival events. We have reconstructed the unitary matrix of the chip and have conducted Bayesian analysis to validate the sampler and confirm that the system exhibits behavior distinct from standard boson sampling. We also built a theoretical description of the system based on the transformation of annihilation operators and, using it, delivered the structure of the transmission matrix and the complexity of our boson sampler in terms of a conventional boson sampler. This work advances photonic quantum computing by demonstrating a resource-efficient method to increase sampling complexity, paving the way for scalable demonstration of quantum advantage with single photons.
\end{abstract}

\maketitle

%\tableofcontents

\input{Main_text/Intro}

\input{Main_text/Theory_short}
\input{Main_text/Experiment}
\input{Main_text/Disc_conc}

\bibliography{bibliography}% Produces the bibliography via BibTeX.
\appendix 
\input{Appendix/Theoretical_description}
\input{Appendix/Demux_chars}
\input{Appendix/Reconstruction}

\end{document}

%% file: Main_text/Intro.tex
\section{Introduction}
Boson sampling is a restricted model of non-universal quantum computation that has garnered significant attention for its potential to demonstrate quantum supremacy with relatively modest resources compared to universal quantum computing \citep{aaronson2013computational, madsen2022quantum, zhong2020quantum}. First proposed by Aaronson and Arkhipov in 2011 \citep{aaronson2010bosonsampling}, boson sampling involves sending indistinguishable photons through a linear optical network and measuring samples from the output distribution, which is proven to be to be computationally intractable for classical computers. 

The basic setup of boson sampling requires single-photon sources, a linear interferometer, and single-photon detectors. Early experiments demonstrated boson sampling with small numbers of photons and modes, such as 2--3 photons in 5--6 mode interferometers \citep{tillmann2013experimental, crespi2013integrated, spring2013boson, broome2013photonic}. Over time, these experiments have scaled up, with more recent implementations involving larger systems, such as 13 modes \citep{spagnolo2014experimental} and 9 modes \citep{wang2015integrated}. Variants of boson sampling have also been proposed to address specific challenges or applications. For instance, scattershot boson sampling uses multiple photon-pair sources to exponentially increase the rate of multi-photon events \citep{wang2015integrated}. Gaussian boson sampling (GBS) employs squeezed states of light instead of single photons, achieving a quantum advantage in a 100$\times$100 optical circuit \citep{wang2020experimental}. Nevertheless, squeezed states of light are sensitive to photon loss, which reduces the complexity of GBS and makes classical simulation with tensor networks possible \cite{oh2024classical}. Meanwhile, the same methods are not applicable to single-photon boson sampling. So, scaling up single-photon boson sampler is a prominent candidate to demonstrate quantum advantage. Moreover, recently new theoretical results relaxed the requirements for the experimental demonstration of quantum computational advantage with single photons, which also stimulates new experimental research \cite{bouland2023complexity, go2024exploring}.  \\
Another notable variant is time-bin encoded boson sampling \cite{he2017time}, which allows temporal correlations among photons and helps to increase the number of modes of the interferometer without blowing up the size of the experimental setup. This approach leverages the interference of photons across multiple time bins, effectively upscaling the number of modes without physically adding more components into the interferometer. This makes it particularly promising for  reaching the quantum advantage threshold, as it can achieve higher complexity with existing hardware. \\
In this work, we present an experimental demonstration of the variation of time-bin boson sampling which we call \textit{loopback boson sampling}. By connecting several output channels with input channels, we create a time-bin encoded system that exhibits behavior distinct from standard boson sampling with the same multimode interferometer, as validated through Bayesian inference. Our setup achieves high photon indistinguishability and demonstrates the feasibility of using optical feedback to enhance the complexity of boson sampling. Our work is inspired by \cite{gao2022quantum} and deepens the analysis of the system, as well as provides instruments for the proper validation of its complexity. In this work we focus on the experimental realisation of loopback boson sampling with simple theoretical model. More comprehensive approach can be found in \cite{biriukovlooptheory2025}.\\
This work shows the potential path towards scaling up single-photon boson sampling without significant increase of physical resources: size of the interferometer, number of detectors, and number of single photons do not increase since we use temporal degree of freedom. It builds on the growing body of research on photonic boson sampling, including recent advances in experimental implementations \citep{wang2015integrated, wang2020experimental} and theoretical studies on the complexity of related models \citep{harrow2017quantum}. 

The rest of this paper is organized as follows. In Section \ref{sec:theor_frame} we briefly describe the theoretical framework of spacio-temporal modes transformation, in Section \ref{section:Setup} we describe the experimental setup and its characteristics. Section \ref{sec:valid} presents the results, including the validation of our boson sampling system. Finally, in Section \ref{sec:conc}, we discuss the implications of our findings and future directions. Details about the theoretical framework and characterization of the experimental setup can be found in the appendix.

%% file: Main_text/Theory_short.tex
\section{Theoretical framework}
\label{sec:theor_frame}
We consider $N$ single photons injected periodically into an $M$-mode interferometer described by a unitary matrix $U$. A subset of $L$ output modes is looped back to $L$ input modes with a delay matching the photon injection period. The system evolution at time step $t$ can be written as
\begin{align}
\hat{b}^{(t)} &= U \hat{a}^{(t)}, \label{eq:1}
\end{align}
where $\hat{a}^{(t)}$ and $\hat{b}^{(t)}$ are vectors of annihilation operators for the input and output modes, respectively.
The feedback condition couples successive temporal iterations as
\begin{align}
\hat{a}^{(t)}_{\text{loop}} = \hat{b}^{(t-1)}_{\text{loop}}.
\end{align}
Iterating this relation yields a block-Toeplitz transformation matrix $U_{\text{total}}$ that describes the evolution across $T$ temporal steps:
\begin{align}
\hat{b}_{\text{total}} = U_{\text{total}} \hat{a}_{\text{total}}.
\end{align}
$U_{\text{total}}$ is a block lower-triangular Toeplitz matrix, which is related to the fact that photons from the iteration $i+1$ cannot be detected in the previous $(M-L)i$ spacio-temporal modes. The effective size of the interferometer grows linearly with the number of temporal iterations $T$, and its complexity is equivalent to that of a conventional boson sampler with $NT$ photons in $(M-L)T+L$ modes. This scaling makes optical feedback a powerful resource for enhancing sampling complexity without increasing the number of physical components.

%% file: Main_text/Experiment.tex
\section{Experimental results}
The central idea of our experiment is to demonstrate that introducing optical feedback loops into a photonic interferometer increases the effective complexity of boson sampling without adding physical resources. 
By reconnecting several output channels to the corresponding inputs, we create temporal correlations between subsequent photon injections. 
These correlations expand the accessible Hilbert space and make the output statistics more complex than in conventional boson sampling. 
\label{sec::experiment}
\subsection{Experimental Setup}
\label{section:Setup}
\begin{figure*}[ht]
    \centering
    \includegraphics*[width=.9\linewidth]{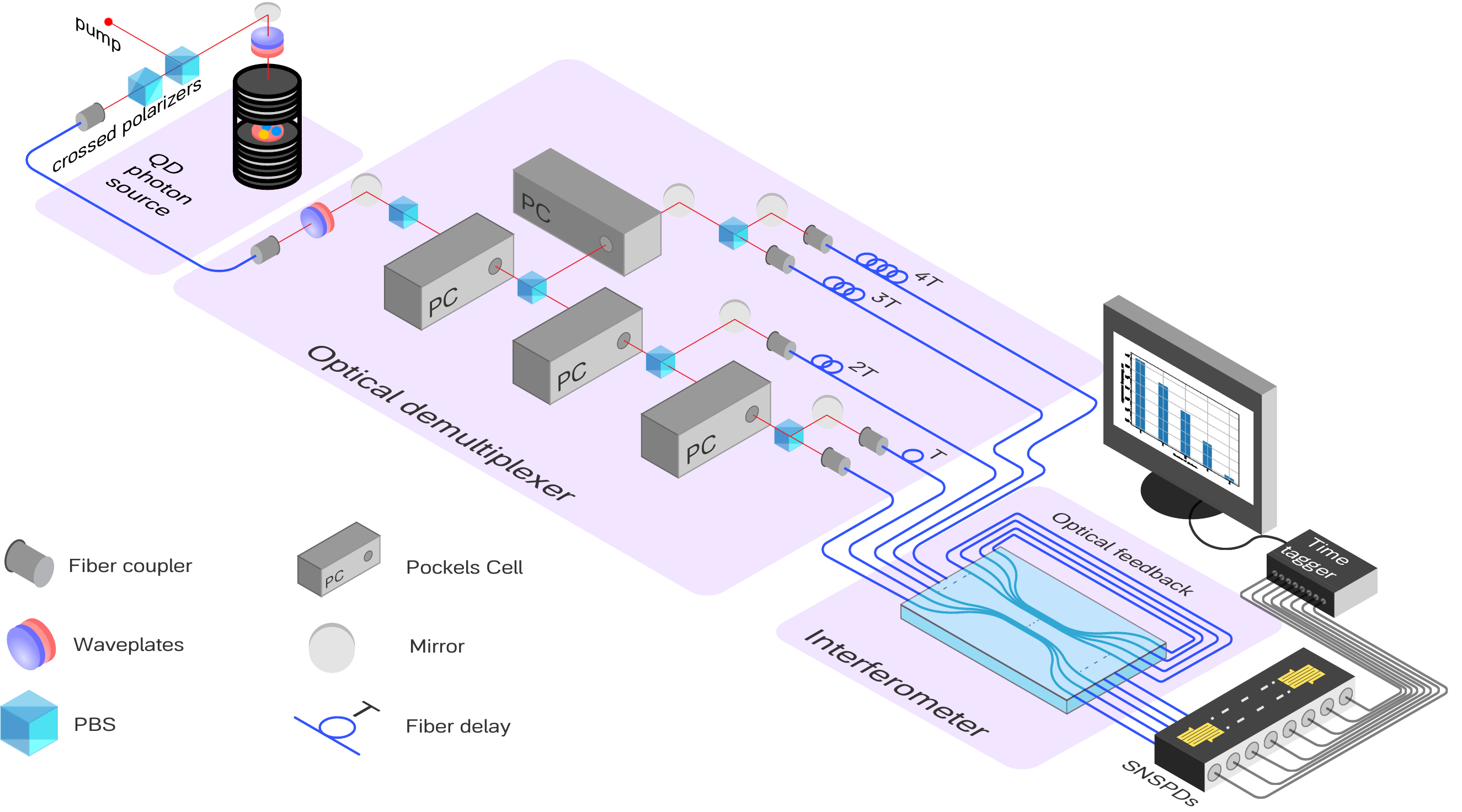}
    \caption{Scheme of the experimental setup. A single-photon source based on an InAs/GaAs quantum dot (QD) in a micropillar resonator emits photons at 926~nm with period $T = 12.1$~ns. A tree-structured demultiplexer composed of four Pockels cells (PCs) and fiber delays prepares five synchronized photons, which are injected into a 25-mode photonic chip with five optical feedback lines. Output photons are detected by superconducting nanowire single-photon detectors (SNSPDs), and detection events are time-tagged for analysis.}
    \label{fig:Setup}
\end{figure*}

The experimental setup is shown in Fig.~\ref{fig:Setup}. 
Five single photons at the chip input are provided by a demultiplexed source based on a resonantly driven InAs/GaAs QD in a micropillar cavity~\cite{RAKHLIN2023}. The emitter, held at 11.9~K in a cryostat, is excited by 320~MHz femtosecond Ti:Sa laser pulses under cross-polarization conditions. The resulting single-photon stream is split into five channels by an electro-optic demultiplexer consisting of four Pockels cells and polarizing beam splitters. An FPGA synchronized with the laser generates control pulses at 5~MHz, routing sequences of 16 consecutive photons (330.4 ~MHz rate) into each channel. Photons are coupled into polarization-maintaining fibers and delayed to achieve temporal synchronization without active stabilization.

The 25-mode interferometer is fabricated using femtosecond laser writing (FSLW) ~\cite{skryabin2024}. The total transmission though the chip is approximately 30~\%, including propagation loss and loss at the interfaces between optical fiber and integrated waveguides. Five outputs are connected to corresponding inputs via fiber feedback lines. The total length of  feedback lines was designed to match the 82.6 MHz repetition rate of the incoming photons, ensuring that only three temporal iterations could occur, wherein photons from preceding iterations might interfere with newly arriving photons. For instance, for 16 incoming photons only the subset of photons \{1,5,9,13\}, \{2,6,10,14\}, \{3,7,11,15\}, \{4,8,12,16\} would interfere.\\
The choice of loopedback outputs was governed by the transmission matrix of the interferometer(see Appendix \ref{sec:reconstruction}): for chosen inputs 9, 10, 13, 16, 17 maximal transmission was into output modes 17, 16, 12, 9, 10 respectively. Those output modes were connected with input channels 5, 19, 8, 12, 22. No active phase stabilization of feedback lines was applied; therefore, theoretical predictions were averaged over random phases. Single photons are coupled to and from the chip via PM-fiber arrays aligned on nanopositioning stages. The experimentally obtained matrix $U_{total}$ was also estimated (see Fig. \ref{fig:matrix}(d) and Appendix \ref{sec:reconstruction} for details) 
Twenty remaining outputs are connected to SNSPDs (Scontel) with 70\% detection efficiency and 10~ns dead time. A time-tagging module (Swabian Instruments) registers photon arrivals, allowing the separation of temporal iterations and reconstruction of photon-count distributions for each cycle.

\subsection{Boson sampling and its validation}
\label{sec:valid}
The goal of this section is to validate that our looped boson sampler operates as a genuine quantum device and that the observed output distributions cannot be reproduced by classical models and conventional boson samplers. 
To achieve this, we first construct and verify a faithful numerical model of the system, then use it to perform Bayesian validation against classical hypotheses and to analyze the effect of optical feedback. 

\begin{figure*}[ht]
    \centering
    \includegraphics*[width=1.\linewidth]{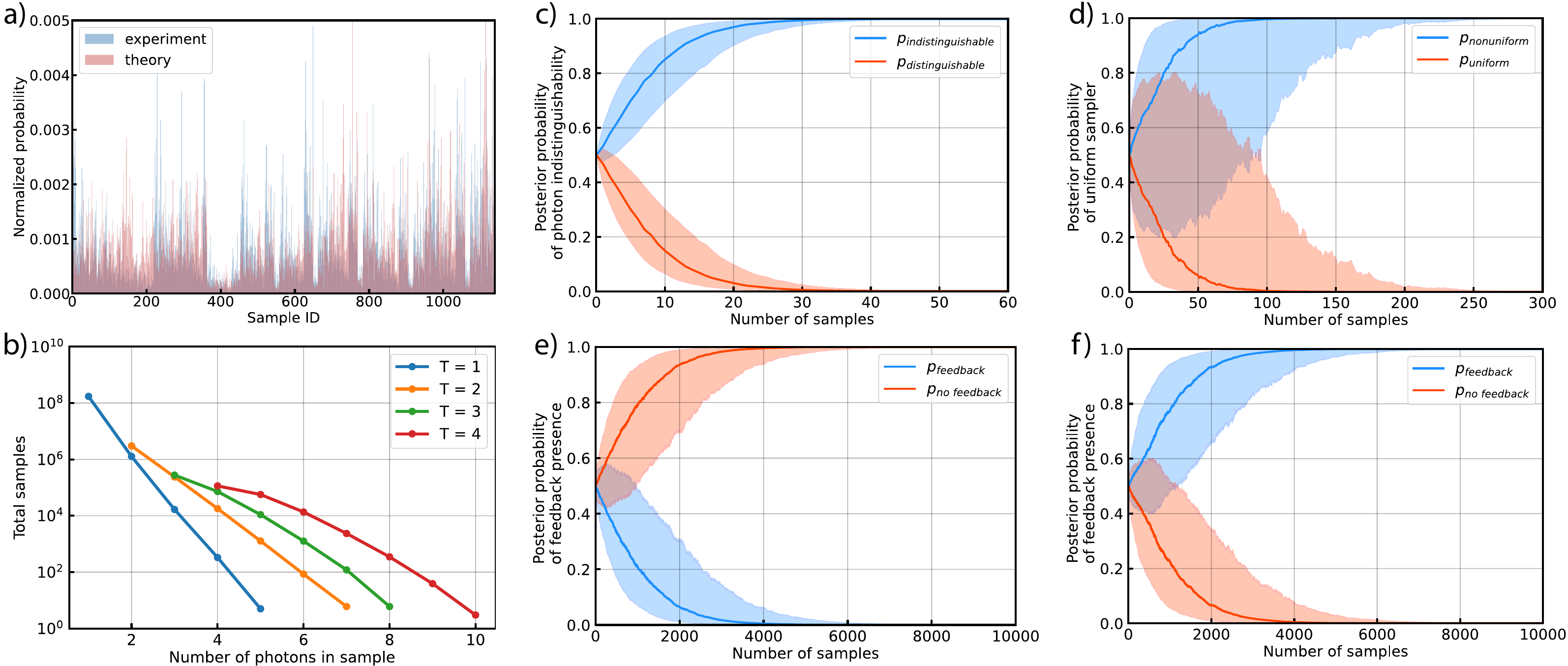}
    \caption{(a) Comparison of experimental and theoretical 3-photon events distribution on the first time iteration. (b) The distribution of the number of detected photons for various number of considered consecutive time-bins. (b-e) Validation of boson sampling device with bayesian inference. Two different colors depict the probability of one of two hypotheses after processing of several samples. Solid lines are mean value of probability by averaged over 1000 different sets of samples, while shaded areas are $90\%$ confidence interval obtained for these 1000 sets of samples. (b) Validation against distinguishable sampler. (c) Validation against uniform sampler. (d) Validation against loop sampler in case of blocked feedback lines. (e) Validation against conventional sampler in case of working feedback lines.}
    \label{fig:validators}
\end{figure*}

A reliable numerical model requires accurate knowledge of the experimental parameters: the interferometer transmission matrix, input and output losses, and single-photon source characteristics (see Appendix~\ref{sec:reconstruction} and \ref{sec:demux_chars}). 
Using this data, we built a numerical model of our device with optical feedback based on the open-source photonic quantum computing framework \verb|Perceval|~\cite{heurtel2023perceval}. 
All known imperfections were included. 
Because the computational cost of sampling increases as that of convensional boson sampler with $NT$ photons in $(M-L)T+L$ modes (see Section~\ref{sec::theory}), we restricted calculations to the first two temporal iterations, which are sufficient for validation.\\
To verify the adequacy of the model, we compared the simulated and experimentally measured photon-count distributions. 
Figure~\ref{fig:validators}(a) shows the marginal three-photon distribution for the first iteration: the theoretical and experimental results agree within statistical error. 
This confirms that the constructed numerical model correctly describes our system and can be used for further validation.\\
Having established a reliable model, we used it to test the quantum behavior of the sampler through Bayesian inference~\cite{bentivegna2015bayesian}. 
Three reference distributions were numerically generated for the first iteration: (i) indistinguishable photons, (ii) distinguishable photons, and (iii) a uniform sampler. 
The corresponding validators are presented in Figs.~\ref{fig:validators}(c,d). 
Only a few dozen samples were sufficient to distinguish the experimental data from the distinguishable sampler, demonstrating that photons are highly indistinguishable and interfere within the chip. Approximately one hundred samples were enough to reject the uniform distribution hypothesis.
Next, we used the same model to investigate the role of optical feedback. If feedback had no effect, the output distributions of consecutive temporal iterations would coincide, corresponding to an ordinary boson sampler. In contrast, statistically distinct distributions were observed experimentally, confirming the influence of feedback. 
We simulated photon distributions in the 20 non-looped modes for both looped and non-looped configurations, averaging over random feedback phases to reflect the absence of active stabilization. 
Experimental data with blocked and open feedback channels are shown in Figs.~\ref{fig:validators}(e,f). 
In both cases, Bayesian validation correctly identified the presence or absence of feedback, confirming that the looped sampler exhibits increased complexity relative to a conventional device.\\
Finally, we compared the effective Hilbert-space size of the looped boson sampler with that of a conventional one. We detected 10-photon events across four consecutive temporal iterations (see Fig. \ref{fig:validators}(b)), corresponding to an effective Hilbert space of approximately 43 qubits. 

%% file: Main_text/Disc_conc.tex
\section{Discussion}
\label{sec:conc}
Our results demonstrate that introducing optical feedback into a linear optical interferometer provides an efficient means of increasing boson-sampling complexity without expanding the physical system. 
By connecting several output channels back to the inputs, we effectively add a temporal degree of freedom, creating correlations between successive photon injections and expanding the accessible Hilbert space. 
The agreement between experiment and theory confirms that this feedback-induced temporal structure is faithfully captured by our model, validating looped boson sampling as a distinct and more complex regime of photonic quantum computation.\\
At the same time, the present architecture has several practical limitations. Even though the effective dimension of the Hilbert state is significantly larger than that of five photons in 25 modes— the same interferometer without feedback— it is still below the threshold for quantum advantage, estimated to require about 50 photons in 120 modes \cite{bouland2023complexity, lund2017quantum}. Because the interferometer’s effective size scales linearly with the number of temporal iterations T, our setup would need six consecutive detection events to reach the 120-mode regime, which is theoretically possible but currently limited by photon loss in the interfaces between optical fiber and integrated photonic circuit, detectors’ efficiency and the length of feedback lines. The loop length, matched to the laser repetition period, restricts us to four temporal iterations, constraining the effective size of the interferometer. This limitation can be overcome if we change the length of the feedback line to match the period between consecutive 16-photon packages routed to the same channel. \\
Photon losses in both the feedback fibers, the femtosecond-laser-written (FSLW) chip and detector inefficiencies significantly reduce multi-photon event rates and limit achievable photon numbers. \\
Additionally, while Bayesian validation successfully distinguishes quantum and classical regimes, its computational cost scales unfavorably with the number of modes and photons, motivating the development of more scalable validation techniques.\\
These challenges suggest several directions for improvement. Reducing propagation and coupling losses in the photonic chip, for example through optimized multiscan fabrication~\cite{tan2022effectively}, would increase multiphoton detection rate.\\
Higher-efficiency single-photon sources~\cite{esmann2024solid} and photon-number-resolving detectors ~\cite{ding2025photon} would allow operation at larger photon numbers. Alternative statistical validation approaches~\cite{agresti2019pattern} could provide faster feedback on device performance for future large-scale implementations. \\
Beyond complexity verification, feedback-enhanced photonic circuits may also find applications in quantum simulations of systems with inherent temporal correlations, such as non-Markovian dynamics or feedback-controlled quantum networks.

\section{Conclusion}
We have experimentally demonstrated boson sampling with optical feedback loops, where temporal correlations among photons enhance the effective sampling complexity without increasing physical resources. Our measurements agree with theoretical predictions and confirm that the looped architecture exhibits behavior distinct from standard boson sampling. This approach establishes a scalable and resource-efficient route toward more complex photonic quantum processors and provides a foundation for future experiments exploiting temporal feedback as a computational resource.

\section{Acknowledgment}
The work was supported by Russian Science Foundation grant 22-12-00353-$\Pi$ (https://rscf.ru/en/project/22-12-00353/). The work was also supported by Rosatom in the framework of the Roadmap for Quantum computing (Contract No. 868-1.3-15/15-2021 dated October 5, 2021) in parts of Quantum Dot fabrication (Contract No. R2152 dated November 19, 2021) and demultiplexed single photon source development, photonic chip fabrication and calibration (Contract No.P2154 dated November 24, 2021). S.P.K. acknowledges support by Ministry of Science and Higher Education of the Russian Federation and South Ural State University (agreement №075-15-2022-1116). Yu.A. B is grateful to the Russian Foundation for the Advancement of Theoretical Physics and Mathematics (BASIS) (Projects №24-2-10-57-1).

%% file: Appendix/Theoretical_description.tex
\section{Theoretical Model of Looped Boson Sampling}
\label{sec::theory}
\subsection{System overview}
\begin{figure}[ht]
    \centering
    \includegraphics*[width=1.\linewidth]{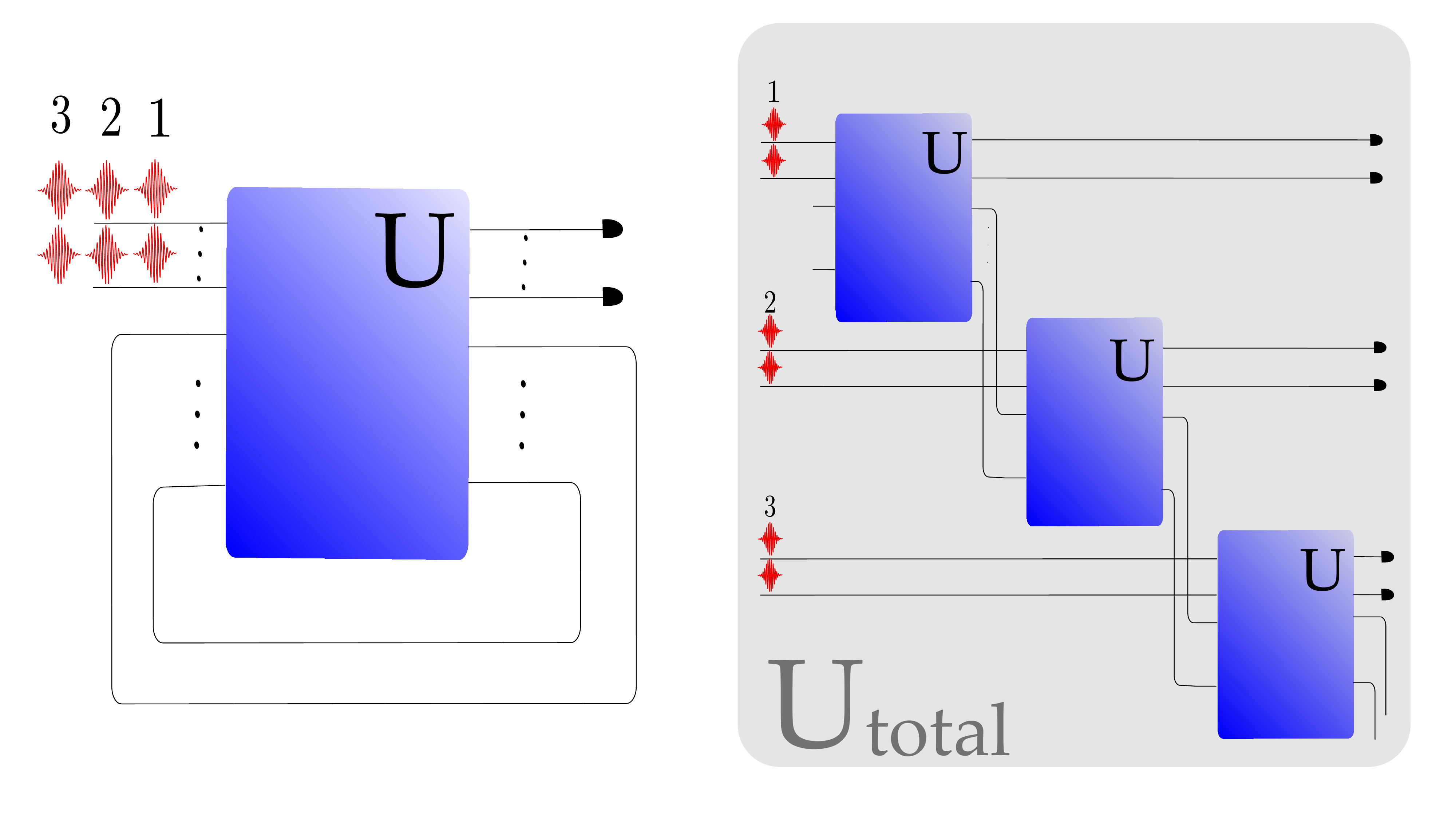}
    \caption{(Left) Scheme of general boson sampler with optical feedback lines with interferometer transmission matrix $U$. (Right) The same interferometer, but with all spacio-temporal modes depicted to consider the system as conventional boson sampler with transfer matrix $U_{total}$. }
    \label{fig:interferometer_mapping}
\end{figure}

We consider a modified boson sampling setup with three key components (see  Fig.\ref{fig:interferometer_mapping}):
\begin{enumerate}
    \item \textbf{Input modes}: \( N \) single photons are injected into an \( M \)-mode interferometer periodically at times \( t = 1, 2, \dots, T \).
    \item \textbf{Interferometer}: A linear-optical unitary transformation \( U \in \mathbb{C}^{M \times M} \) is applied at each time step.
    \item \textbf{Output Modes}: The last \( L \) output modes are looped back to the last \( L \) input modes with a delay matching the photon injection interval. We call this \textit{optical feedback channels}. We assume that the optical feedback channels makes photons from iteration \( t-1 \) arrive at the input simultaneously with new photons at iteration \( t \). The first $M-L$ output modes end up by single photon detection.  
\end{enumerate}

Let's introduce necessary notation:
\begin{itemize}
    \item \( \hat{a}_j^{(t)} \) -- an annihilation operator for input mode \( j \) at time \( t \); 
    \item  \( \hat{b}_j^{(t)} \) -- an annihilation operator for output mode \( j \) at time \( t \); 
    \item \(\hat{\mathbf{a}}^{(t)} = [\hat{a}_1^{(t)}, \dots, \hat{a}_M^{(t)}]^T \) -- an input vector at time \( t \);
    \item \(\hat{\mathbf{b}}^{(t)} = [\hat{b}_1^{(t)}, \dots, \hat{b}_M^{(t)}]^T \) -- output vector at time \( t \);
    \item  \( \hat{\mathbf{a}}_{\text{ext}}^{(t)} \)-- external photons injected at time \( t \) (first \( M-L \) modes);
    \item  \( \hat{\mathbf{a}}_{\text{loop}}^{(t)} \)-- looped photons (last \( L \) modes).
\end{itemize}
We suppose, without loss of generality, that the looped modes are the last $L$ modes and that $M-L+i$  looped mode is coupled to $M-L+i$ input mode, so the coupling relation can be written as follows:
\[
\hat{\mathbf{a}}_{\text{loop}}^{(t)} = 
\begin{cases} 
\mathbf{0}, & t = 1 \\
\text{(Last } L \text{ components of } \hat{\mathbf{b}}^{(t-1)}), & t \geq 2.
\end{cases}.
\]
If there are non-compensated phase shifts $\phi_L$ in optical feedback lines, they can be incorporated in the unitary transformation. At each time step, the interferometer acts as:
\[
\hat{\mathbf{b}}^{(t)} = U \hat{\mathbf{a}}^{(t)}, \quad U^\dagger U = I_M.
\]
We partition \( U \) into submatrices for the non-looped and looped modes:
\[
U = \begin{pmatrix}
U_{\text{EE}} & U_{\text{EL}} \\
U_{\text{LE}} & U_{\text{LL}}
\end{pmatrix},
\]
where:
\begin{itemize}
    \item \( U_{\text{EE}} \in \mathbb{C}^{(M-L) \times (M-L)} \) maps non-looped inputs to non-looped outputs (subscript E for \textit{external});
    \item  \( U_{\text{EL}} \in \mathbb{C}^{L \times (M-L)} \) maps non-looped inputs to looped outputs (subscript L for \textit{looped}).
    \item \( U_{\text{EL}} \in \mathbb{C}^{L \times (M-L)} \) maps looped inputs to non-looped outputs.
    \item \( U_{\text{loop}} \in \mathbb{C}^{L \times L} \) maps looped inputs to looped outputs.
\end{itemize}

\subsection{Output distribution}
To obtain the transformation matrix for the input vectors $\hat{\mathbf{a}}^{(t)}$ we can write recurrent equations for both looped and non-looped output modes. Substituting \( \hat{\mathbf{b}}^{(t)} = U \hat{\mathbf{a}}^{(t)} \), we get:
\begin{equation}
\begin{cases} 
 \hat{\mathbf{b}}^{(t)}_{ext}=U_{EE}\hat{\mathbf{a}}_{\text{ext}}^{(t)} + U_{EL}\hat{\mathbf{a}}_{\text{loop}}^{(t)}\\
 
 \hat{\mathbf{a}}_{\text{loop}}^{(t+1)} = \hat{\mathbf{b}}_{\text{loop}}^{(t)} = U_{LE}\hat{\mathbf{a}}_{\text{ext}}^{(t)} + U_{LL}\hat{\mathbf{a}}_{\text{loop}}^{(t)}
 \end{cases}
\end{equation}
Tracing back to $t=1$ we may obtain:
\begin{equation}
\hat{\mathbf{b}}^{(t)}_{ext} = U_{EE}\hat{\mathbf{a}}_{\text{ext}}^{(t)}  + U_{EL}\sum_{k=1}^{t-1}[U_{LL}]^{t-k-1}U_{LE}\hat{\mathbf{a}}_{\text{ext}}^{(k)} 
\end{equation}
So if we introduce vectors of all spatio-temporal accessible input and output modes:
\begin{align}
\hat{\mathbf{a}}_{total} = [\hat{\mathbf{a}}_{ext}^{(1)}\dots\hat{\mathbf{a}}_{ext}^{(T)}]^T,\\
\hat{\mathbf{b}}_{total} = [\hat{\mathbf{b}}_{ext}^{(1)}\dots\hat{\mathbf{b}}_{ext}^{(T)}]^T,
\end{align}
then the total unitary $U_{total}$ will have the following block structure:
\begin{equation}
[U_{total}]_{tt'} = \begin{cases}
    U_{EE}, \text{if } t = t',\\
    U_{EL}[U_{LL}]^{t-t'-1}U_{LE},  \text{if }t>t',\\
    0,  \text{if } t<t'.
\end{cases}
\end{equation}
For example, for the case of $T=3$ we have:
\begin{equation}
U_{\text{total}} = \begin{pmatrix}
 U_{EE}& 0 & 0 \\
U_{EL}U_{LE} &  U_{EE} & 0 \\
U_{EL}U_{LL}U_{LE}&U_{EL}U_{LE}& U_{EE}
\end{pmatrix},
\end{equation}
which is a block lower-triangular Toeplitz matrix. This matrix is far from random Haar matrix, but is much more rich than just block-diagonal matrix describing the case with no optical feedback lines. \\
It is easy to see that this matrix is not unitary since we consider only subset of modes at each iteration. This matrix can only be used to calculate the probabilities of events when all the $NT$ photons are detected. \\
To get access to the case where not all photons are detected we can use the following hint.
If we suppose that at last iteration $T$ all looped channels are unlooped and detected. It adds $L$ modes to $\hat{\mathbf{b}}_{total}$  output vector, so it becomes:
\begin{equation}
    \hat{\mathbf{b}}_{total}' = [\hat{\mathbf{b}}_{ext}^{(1)}\dots\hat{\mathbf{b}}_{ext}^{(T)}\hat{\mathbf{b}}_{loop}^{(T)}]^T
\end{equation}
This procedure also adds new lines into $U_{total}$ so it becomes rectangular with shape $[(M-L)T+L]\times (M-L)T$. The block structure of this new matrix $U_{total}'$ is:
\begin{equation}
[U_{total}]_{tt'} = \begin{cases}
    [U_{total}]_{tt'}, \text{for first } (M-L)T \text{ rows and columns},  \\
    [U_{LL}]^{t-t'}U_{LE}, t = T, \text{for last } L \text{ rows and columns}\\
\end{cases}
\end{equation}
If, for instance, $T=3$: 
\begin{equation}
U'_{\text{total}} = \begin{pmatrix}
 U_{EE}& 0 & 0 \\
U_{EL}U_{LE} &  U_{EE} & 0 \\
U_{EL}U_{LL}U_{LE}&U_{EL}U_{LE}& U_{EE}\\
U_{LL}U_{LL}U_{LE}&U_{LL}U_{LE}& U_{LE}
\end{pmatrix},
\end{equation}
Now there are no unconsidered spacio-temporal modes, so the probability distribution, obtained by calculation of permanents of submatrices of $U_{total}'$ with size $NT\times NT$,  is normalized. \\
Actually, we can't afford to unloop the channels so fast. It means that we don't have access to the last looped modes $\hat{\mathbf{b}}_{loop}^{(T)}$. To deal with it, we only need to trace out those last modes in the obtained probability distribution. The method proposed above shows that the complexity of boson sampling setup with optical feedback grows with the number of considered time iterations $T$ and is equivalent to boson sampler with $NT$ photons in $(M-L)T+L$ modes.

\subsection{Boson sampling validation}
\label{subsec::bayes}
The crucial part of the boson sampling experiments is validation of the correct work of the device. Many different techniques have been proposed to distinguish a true boson sampler from pseudo-samplers that may obey similarly to the true one but are easy to simulate with classical computer (see \cite{Li:18} for detailed review). \\
To validate our loopback sampler, it is necessary to observe the statistical difference between probability distributions in non-looped modes with and without optical feedback. Unfortunately, due to photon loss it may be necessary to accumulate statistics for a long amount of time to see significant difference between those two distributions. But if we have access to probability distributions at each temporal iteration, it is enough to compare the difference between the first and second iterations to see whether there is an effect of optical feedback or not. We have adapted the Bayesian inference approach from \cite{bentivegna2015bayesian} to distinguish between looped and non-looped boson samplers. The procedure is as follows.\\
We have two samplers: $A$ -- boson sampling device has optical feedback; $B$ --  boson sampling device without optical feedback. Initially, we don't know what device we have, so the probabilities of both hypotheses are equal: $P(A)=P(B) = 0.5$. Then we start acquiring experimental samples. For each sample $n_1$ we can calculate the probability that it appears from sampler $A$ -- $(P(n_1|A)$ and from sampler $B$ -- $P(n_1|B)$. The total probability for sample $n_1$ to appear is $P(n_1) =  P(n_1|B)P(B) + P(n_1|A)P(A) $ Now we can update the probabilities of having one of two samplers with information about the sample $n_1$ using the Bayes theorem:
\begin{align}
P(A|n_1) = \frac{P(n_1|A)P(A)}{P(n_1)}, \\
P(B|n_1) = \frac{P(n_1|B)P(B)}{P(n_1)}.
\end{align}
Now using Bayesian inference we make $P(A)=P(A|n_1), P(B) = P(B|n_1)$ and when we get a new sample $n_2$ we will use updated probabilities. Following \cite{bentivegna2015bayesian} we may consider the ratio of $P(A)$ and $P(B)$. So after $N$ samples the ratio will be
\begin{equation}
    \frac{P(A)}{P(B)} = \prod_{i=1}^N\frac{P(n_i|A)}{P(n_i|B)} = \chi.
\end{equation}
To make sure that $P(A)+P(B) = 1$ we introduce a normalization factor $\chi+1$, so $P(A) = \frac{\chi}{\chi+1}$ and $P(B)=\frac{1}{\chi+1}$. If samplers A and B are distinguishable, after some number of samples, one of the probabilities will be close to zero.\\
In the same manner, we will test our sampler against distinguishable and uniform boson sampler hypotheses.

%% file: Appendix/Demux_chars.tex
\section{Optical demultiplexer characterisation}
\label{sec:demux_chars}
\begin{figure*}[ht!]
\centering
\fbox{\includegraphics[width=1\linewidth]{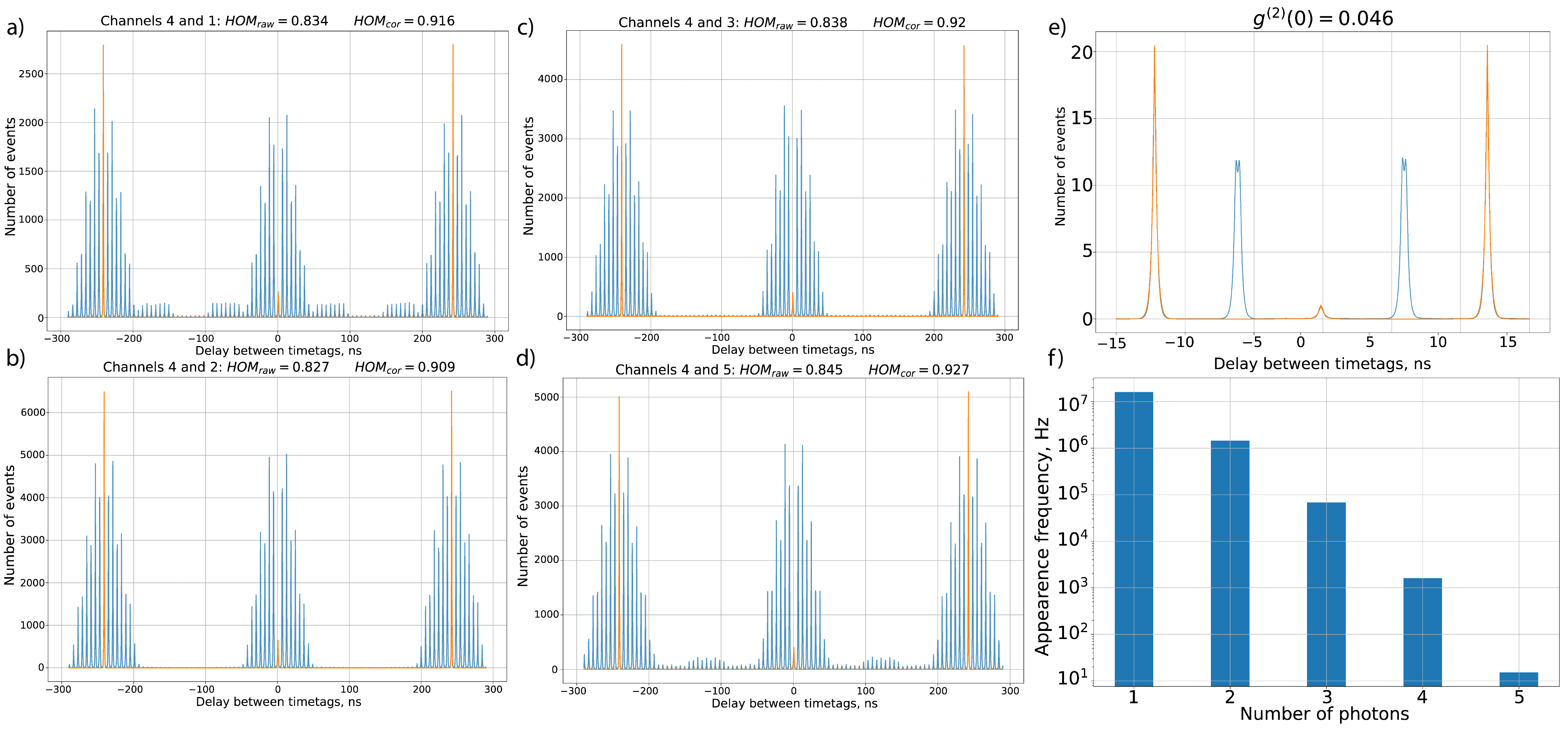}}
\caption{Characteristics of single photon source and spacio-temporal demultiplexer. All the measurements were conducted with pump laser pulse frequency double, so there are sidebands at time delays around $\pm 6$~ns. a-d) Results of measurements of pairwise single photons' indistinguishability. All indistinguishabilities are measures against the 4th channel as a reference. The mean value of indistinguishability is $0.918$. e) Measurement of normalized $g^{(2)}(0)$ showing multiphoton component in single photon radiation. f) Multiphoton events frequencies showing how many photons at the outputs of demultiplexer we observed. These values are not corrected by SNSPDs efficiencies.}
\label{fig::Demux_chars}
\end{figure*}

Single photons with a lifetime of 170 ps at a central wavelength of $926$ nm were coupled to a single-mode fiber with an efficiency of 15.3\%. The purity of single photons was estimated using a conventional Hanbury Brown-Twiss type measurement \cite{HANBURYBROWN1956}, which yielded  $g^{(2)}(0) = 0.046$ (see Fig. \ref{fig::Demux_chars}e).\\
The indistinguishability of single photons from different demultiplexer channels was determined using a Hong-Ou-Mandel (HOM) interferometer \cite{HOM87} and to account for experimental imperfections (asymmetric beam-splitter, non-uniform losses in two input channels of the beam-splitter, non-zero $g^{(2)}(0)$) we used the correction formula from \cite{biriukov2025} since it was obtained specially for the demultiplexed photon source. An average HOM interference visibility value of $0.918$ was obtained (see \ref{fig::Demux_chars} a-d). The measured five-photon generation rate at the outputs of the demultiplexer is 15 Hz with an average channel transmission efficiency of $80\%$ (see Fig. \ref{fig::Demux_chars} f).\\

%% file: Appendix/Reconstruction.tex
\section{Unitary matrix reconstruction}
\label{sec:reconstruction}
\begin{figure*}[ht!]
    \centering
    \includegraphics*[width=1\linewidth]{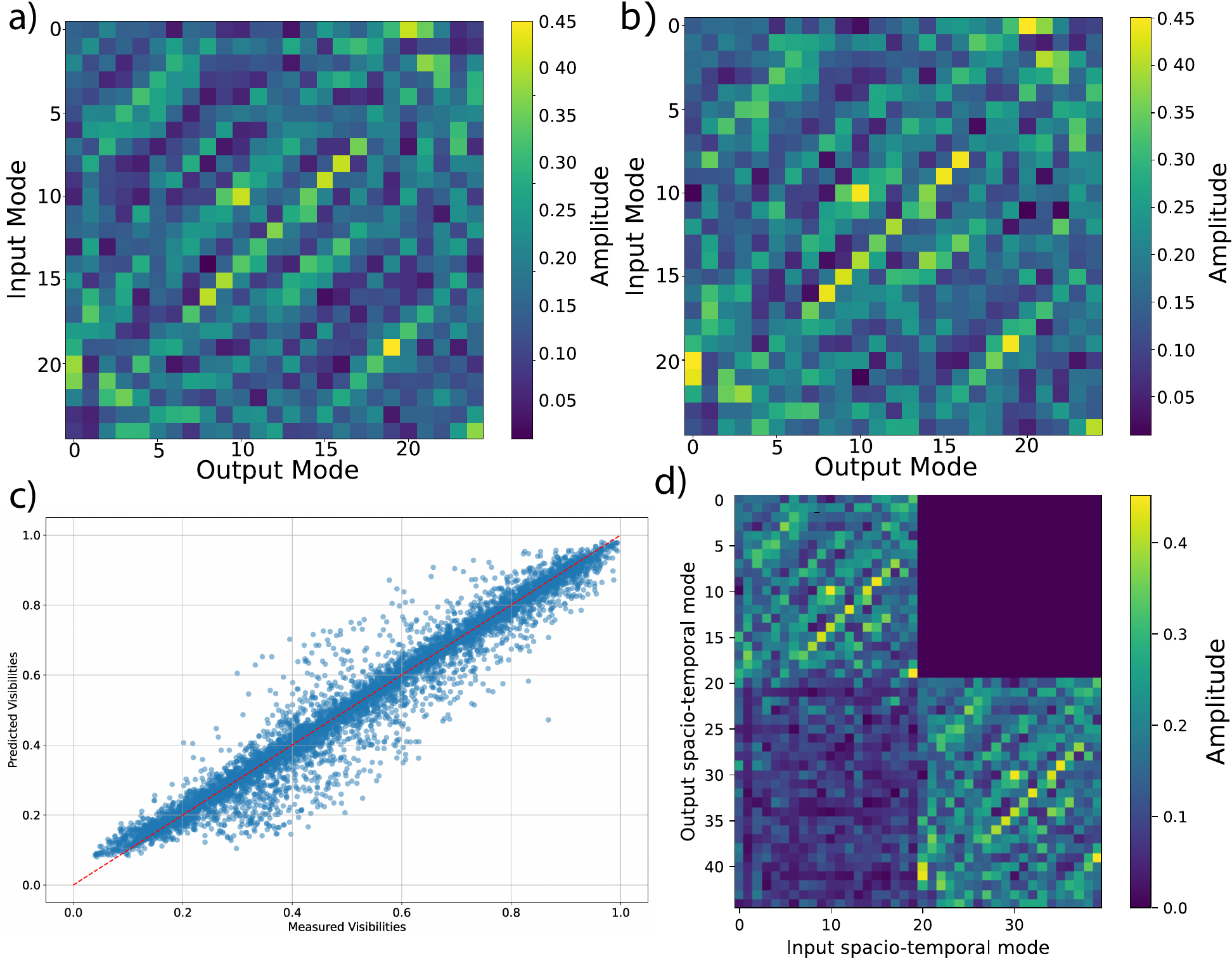}
    \caption{Transmission matrix reconstruction results. (a) Experimentally measured matrix modules of transmission matrix. (b) Matrix of modules obtained after optimization procedure. Fidelity between experimental and numerical matrices is $98.7\%$ (c) Comparison between two-photon interference visibilities obtained via cross-correlation measurements and after optimization procedure. Mean absolute error is $0.033$. (d) Experimentally reconstructed transmission matrix $U_{total}$ obtained for 2 consequent temporal iterations.}
    \label{fig:matrix}
\end{figure*}

To use the Bayesian validation procedure described in \ref{subsec::bayes} knowledge of the interferometer transmission matrix is required. For this purpose, we used the unitary matrix tomography procedure from \cite{biriukov2025}.  We measured modules of transmission matrix by consecutively injecting single photons at all $M=25$ ports of the interferometer and by measuring single photon counts at all outputs. This procedure also allowed us to obtain estimates for non-uniform losses at the inputs and at the outputs of the interferometer (see \cite{biriukov2025} for details), which are necessary for simulation of the boson sampling circuit. \\
After measuring modules of the transmission matrix, we injected two single photons simultaneously at different pairs of input channels to get phases of matrix elements by measuring visibilities of two-photon interference captured by cross-correlation histograms between different pairs of output channels. Using a time-tagging module, it was possible to measure all cross-correlations between different pairs of 20 output channels simultaneously, which is $C^2_{25}=300$ (20 is the number of available single-photon detectors). By varying about $2M$ pairs of input channels we measured over $14 000$ two-photon visibilities and then started optimization procedure to find optimal interferometer matrix, input losses, and source parameters that fit both measured modules and visibilities. Since all visibilities can be expressed directly through elements of matrix (see \cite{biriukov2025}) it is possible to write gradient optimizer using neural-network libraries like \verb|torch|.  This approach is more numerically stable and has better performance over non-gradient approaches.\\
The results of optimization are shown on Fig. \ref{fig:matrix}. 
Source parameters found by optimizer were the same as found independently in Section \ref{section:Setup}. Mean absolute error between all predicted and measured visibilities was $0.052$, while fidelity between matrices of predicted and measured modules, defined as
\begin{equation}
F(M_{e}, M_{t}) = \frac{|Tr(M_{e}^{\dagger}M_{t})|^2}{Tr(M_{e}^{\dagger}M_{e})Tr(M_{t}^{\dagger}M_{t})}.
\label{Fid_eq}
\end{equation}
was $98.7\%$. These results make us believe that the reconstructed matrix correspond to the transmission matrix of interferometer. To our knowledge, it is the first experimental reconstruction of the unitary matrix of such dimension.

%% file: bibliography.bib
@article{zhong2020quantum,
  title={Quantum computational advantage using photons},
  author={Zhong, Han-Sen and Wang, Hui and Deng, Yu-Hao and Chen, Ming-Cheng and Peng, Li-Chao and Luo, Yi-Han and Qin, Jian and Wu, Dian and Ding, Xing and Hu, Yi and others},
  journal={Science},
  volume={370},
  number={6523},
  pages={1460--1463},
  year={2020},
  publisher={American Association for the Advancement of Science}
}

@article{gao2022quantum,
  title={Quantum advantage with membosonsampling},
  author={Gao, Jun and Wang, Xiao-Wei and Zhou, Wen-Hao and Jiao, Zhi-Qiang and Ren, Ruo-Jing and Fu, Yu-Xuan and Qiao, Lu-Feng and Xu, Xiao-Yun and Zhang, Chao-Ni and Pang, Xiao-Ling and others},
  journal={Chip},
  volume={1},
  number={2},
  pages={100007},
  year={2022},
  publisher={Elsevier}
}

@article{madsen2022quantum,
  title={Quantum computational advantage with a programmable photonic processor},
  author={Madsen, Lars S and Laudenbach, Fabian and Askarani, Mohsen Falamarzi and Rortais, Fabien and Vincent, Trevor and Bulmer, Jacob FF and Miatto, Filippo M and Neuhaus, Leonhard and Helt, Lukas G and Collins, Matthew J and others},
  journal={Nature},
  volume={606},
  number={7912},
  pages={75--81},
  year={2022},
  publisher={Nature Publishing Group UK London}
}

@article{Li:18,
author = {Daniel J. Brod and Ernesto F. Galvão and Andrea Crespi and Roberto Osellame and Nicolò Spagnolo and Fabio Sciarrino},
journal = {Advanced Photonics},
number = {11},
pages = {034001},
publisher = {},
title = {Photonic implementation of boson sampling: a review},
volume = {1},
month = {3},
year = {2019},
url = {https://m.researching.cn/articles/OJ1d41e4f96f414592},
}

@article{biriukovlooptheory2025,
  title={Simulation of boson sampling this optical feedback},
   author={Scheel, Stefan},
  journal={arXiv preprint quant-ph/250617},
  year={2025}
}

@article{bentivegna2015bayesian,
  title={Bayesian approach to boson sampling validation},
  author={Bentivegna, Marco and Spagnolo, Nicolo and Vitelli, Chiara and Brod, Daniel J and Crespi, Andrea and Flamini, Fulvio and Ramponi, Roberta and Mataloni, Paolo and Osellame, Roberto and Galv{\~a}o, Ernesto F and others},
  journal={International Journal of Quantum Information},
  volume={12},
  number={07n08},
  pages={1560028},
  year={2015},
  publisher={World Scientific}
}

@article{skryabin2024,
  title = {Femtosecond-laser-written low-loss multiscan waveguides in fused silica},
  author = {Skryabin, N.N. and Zhuravitskii, S.A. and Dyakonov, I.V. and Straupe, S.S. and Kalinkin, A.A. and Kulik, S.P.},
  journal = {Phys. Rev. Appl.},
  volume = {22},
  issue = {6},
  pages = {064079},
  numpages = {13},
  year = {2024},
  month = {Dec},
  publisher = {American Physical Society},
  doi = {10.1103/PhysRevApplied.22.064079},
  url = {https://link.aps.org/doi/10.1103/PhysRevApplied.22.064079}
}

@article{RAKHLIN2023,
title = {Demultiplexed single-photon source with a quantum dot coupled to microresonator},
journal = {Journal of Luminescence},
volume = {253},
pages = {119496},
year = {2023},
issn = {0022-2313},
doi = {https://doi.org/10.1016/j.jlumin.2022.119496},
url = {https://www.sciencedirect.com/science/article/pii/S0022231322007712},
author = {M.V. Rakhlin and A.I. Galimov and I.V. Dyakonov and N.N. Skryabin and G.V. Klimko and M.M. Kulagina and Yu.M. Zadiranov and S.V. Sorokin and I.V. Sedova and Yu.A. Guseva and D.S. Berezina and Yu.M. Serov and N.A. Maleev and A.G. Kuzmenkov and S.I. Troshkov and K.V. Taratorin and A.K. Skalkin and S.S. Straupe and S.P. Kulik and T.V. Shubina and A.A. Toropov},
}

@Article{HANBURYBROWN1956,
author={Hanbury Brown, R.
and Twiss, R. Q.},
title={The Question of Correlation between Photons in Coherent Light Rays},
journal={Nature},
year={1956},
month={Dec},
day={01},
volume={178},
number={4548},
pages={1447-1448},
issn={1476-4687},
doi={10.1038/1781447a0},
url={https://doi.org/10.1038/1781447a0}
}

@article{HOM87,
  title = {Measurement of subpicosecond time intervals between two photons by interference},
  author = {Hong, C. K. and Ou, Z. Y. and Mandel, L.},
  journal = {Phys. Rev. Lett.},
  volume = {59},
  issue = {18},
  pages = {2044--2046},
  numpages = {0},
  year = {1987},
  month = {Nov},
  publisher = {American Physical Society},
  doi = {10.1103/PhysRevLett.59.2044},
  url = {https://link.aps.org/doi/10.1103/PhysRevLett.59.2044}
}

@misc{biriukov2025,
      title={Noise-tolerant tomography of multimode linear optical interferometers with single photons}, 
      author={Yu. A. Biriukov and R. D. Morozov and I. V. Dyakonov and M. V. Rakhlin and A. I. Galimov and G. V. Klimko and S. V. Sorokin and I. V. Sedova and M. M. Kulagina and Yu. M. Zadiranov and A. A. Toropov and A. A. Korneev and S. P. Kulik and S. S. Straupe},
      year={2025},
      eprint={2506.20490},
      archivePrefix={arXiv},
      primaryClass={quant-ph},
      url={https://arxiv.org/abs/2506.20490}, 
}

@article{heurtel2023perceval,
  title={Perceval: A software platform for discrete variable photonic quantum computing},
  author={Heurtel, Nicolas and Fyrillas, Andreas and De Gliniasty, Gr{\'e}goire and Le Bihan, Rapha{\"e}l and Malherbe, S{\'e}bastien and Pailhas, Marceau and Bertasi, Eric and Bourdoncle, Boris and Emeriau, Pierre-Emmanuel and Mezher, Rawad and others},
  journal={Quantum},
  volume={7},
  pages={931},
  year={2023},
  publisher={Verein zur F{\"o}rderung des Open Access Publizierens in den Quantenwissenschaften}
}

@article{aaronson2013computational,
  author = {Aaronson, Scott and Arkhipov, Alex},
  title = {The computational complexity of linear optics},
  journal = {Theory of Computing},
  volume = {9},
  number = {1},
  pages = {143--252},
  year = {2013},
  doi = {10.4086/toc.2013.v009a004}
}

@article{broome2013photonic,
  author = {Broome, M. A. and Fedrizzi, A. and Rahimi-Keshari, S. and Dove, J. and Aaronson, Scott and Ralph, T. C. and White, A. G.},
  title = {Photonic boson sampling in a tunable circuit},
  journal = {Science},
  volume = {339},
  number = {6121},
  pages = {794--798},
  year = {2013},
  doi = {10.1126/science.1231440}
}

@misc{aaronson2010bosonsampling,
  author = {Aaronson, Scott and Arkhipov, Alex},
  title = {{BosonSampling} is hard to classically simulate},
  year = {2010},
  eprint = {1011.3795},
  archivePrefix = {arXiv},
  primaryClass = {quant-ph}
}

@article{spring2013boson,
  author = {Spring, J. B. and Metcalf, B. J. and Humphreys, P. C. and Kolthammer, W. S. and Jin, X.-M. and Barbieri, M. and Datta, A. and Thomas-Peter, N. and Langford, N. K. and Kundys, D. and Gates, J. C. and Smith, B. J. and Smith, P. G. R. and Walmsley, I. A.},
  title = {Boson sampling on a photonic chip},
  journal = {Science},
  volume = {339},
  number = {6121},
  pages = {798--801},
  year = {2013},
  doi = {10.1126/science.1231692}
}

@article{tillmann2013experimental,
  author = {Tillmann, M. and Dakic, B. and Heilmann, R. and Nolte, S. and Szameit, A. and Walther, P.},
  title = {Experimental boson sampling},
  journal = {Nature Photonics},
  volume = {7},
  number = {7},
  pages = {540--544},
  year = {2013},
  doi = {10.1038/nphoton.2013.102}
}

@article{crespi2013integrated,
  author = {Crespi, A. and Osellame, R. and Ramponi, R. and Brod, D. J. and Galvão, E. F. and Spagnolo, N. and Vitelli, C. and Maiorino, E. and Mataloni, P. and Sciarrino, F.},
  title = {Integrated multimode interferometers with arbitrary designs for photonic boson sampling},
  journal = {Nature Photonics},
  volume = {7},
  number = {7},
  pages = {545--549},
  year = {2013},
  doi = {10.1038/nphoton.2013.112}
}

@article{wang2015integrated,
  author = {Wang, C. and Gao, X. and Shi, J. and Wang, Y. and He, Y. and Zhang, R. and Wang, H. and He, Y. and Wang, Y. and Pan, J. and Lu, C.},
  title = {Integrated scattershot boson sampling},
  journal = {Science Advances},
  volume = {1},
  number = {10},
  pages = {e1400255},
  year = {2015},
  doi = {10.1126/sciadv.1400255}
}

@article{wang2020experimental,
  author = {Wang, Z. and Qin, S. and Li, J. and Wang, H. and Wang, Y. and He, Y. and Wang, Y. and Pan, J. and Lu, C.},
  title = {Experimental demonstration of Gaussian boson sampling},
  journal = {Science},
  volume = {370},
  number = {6523},
  pages = {1458--1461},
  year = {2020},
  doi = {10.1126/science.abe2828}
}

@article{spagnolo2014experimental,
  author = {Spagnolo, N. and Vitelli, C. and Bentivegna, M. and Brod, D. J. and Crespi, A. and Flamini, F. and Giacomini, S. and Milani, G. and Ramponi, R. and Mataloni, P. and Osellame, R. and Galvão, E. F. and Sciarrino, F.},
  title = {Experimental validation of photonic boson sampling},
  journal = {Nature Photonics},
  volume = {8},
  number = {9},
  pages = {615--620},
  year = {2014},
  doi = {10.1038/nphoton.2014.135}
}

@article{harrow2017quantum,
  author = {Harrow, A. W. and Montanaro, A.},
  title = {Quantum computational supremacy},
  journal = {Nature},
  volume = {549},
  number = {7671},
  pages = {203--209},
  year = {2017},
  doi = {10.1038/nature23458}
}

@article{he2017time,
  title={Time-bin-encoded boson sampling with a single-photon device},
  author={He, Yu and Ding, X and Su, Z-E and Huang, H-L and Qin, J and Wang, C and Unsleber, S and Chen, C and Wang, H and He, Y-M and others},
  journal={Physical review letters},
  volume={118},
  number={19},
  pages={190501},
  year={2017},
  publisher={APS}
}

@article{lund2017quantum,
  title={Quantum sampling problems, BosonSampling and quantum supremacy},
  author={Lund, Austin P and Bremner, Michael J and Ralph, Timothy C},
  journal={npj Quantum Information},
  volume={3},
  number={1},
  pages={15},
  year={2017},
  publisher={Nature Publishing Group UK London}
}

@article{oh2024classical,
  title={Classical algorithm for simulating experimental Gaussian boson sampling},
  author={Oh, Changhun and Liu, Minzhao and Alexeev, Yuri and Fefferman, Bill and Jiang, Liang},
  journal={Nature Physics},
  volume={20},
  number={9},
  pages={1461--1468},
  year={2024},
  publisher={Nature Publishing Group UK London}
}

@article{bouland2023complexity,
  title={Complexity-theoretic foundations of bosonsampling with a linear number of modes},
  author={Bouland, Adam and Brod, Daniel and Datta, Ishaun and Fefferman, Bill and Grier, Daniel and Hernandez, Felipe and Oszmaniec, Michal},
  journal={arXiv preprint arXiv:2312.00286},
  year={2023}
}

@article{go2024exploring,
  title={Exploring shallow-depth boson sampling: Toward a scalable quantum advantage},
  author={Go, Byeongseon and Oh, Changhun and Jiang, Liang and Jeong, Hyunseok},
  journal={Physical Review A},
  volume={109},
  number={5},
  pages={052613},
  year={2024},
  publisher={APS}
}

@article{esmann2024solid,
  title={Solid-state single-photon sources: recent advances for novel quantum materials},
  author={Esmann, Martin and Wein, Stephen C and Ant{\'o}n-Solanas, Carlos},
  journal={Advanced Functional Materials},
  volume={34},
  number={30},
  pages={2315936},
  year={2024},
  publisher={Wiley Online Library}
}

@article{ding2025photon,
  title={Photon-number-resolving single-photon detector with a system detection efficiency of 98\% and photon-number resolution of 32},
  author={Ding, Chaomeng and Zhang, Xingyu and Xiong, Jiamin and Xiao, You and Zhang, Tianzhu and Huang, Jia and Xu, Hongxin and Liu, Xiaoyu and You, Lixing and Wang, Zhen and others},
  journal={ACS Photonics},
  year={2025},
  publisher={ACS Publications}
}

@article{tan2022effectively,
  title={Effectively writing low propagation and bend loss waveguides in the silica glass by using a femtosecond laser},
  author={Tan, Dezhi and Sun, Xiaoyu and Li, Zengling and Qiu, Jianrong},
  journal={Optics Letters},
  volume={47},
  number={18},
  pages={4766--4769},
  year={2022},
  publisher={Optica Publishing Group}
}

@article{agresti2019pattern,
  title={Pattern recognition techniques for boson sampling validation},
  author={Agresti, Iris and Viggianiello, Niko and Flamini, Fulvio and Spagnolo, Nicol{\`o} and Crespi, Andrea and Osellame, Roberto and Wiebe, Nathan and Sciarrino, Fabio},
  journal={Physical Review X},
  volume={9},
  number={1},
  pages={011013},
  year={2019},
  publisher={APS}
}
